\documentclass[prl,twocolumn,draft,amsmath,showpacs]{revtex4}
\usepackage{graphics}

\begin{document}

\bibliographystyle{prsty}
\input epsf

\title{NMR and dc-susceptibility studies of NaVGe$_{2}$O$_{6}$}

\author{B. Pedrini$^{1}$, J. L. Gavilano$^{1}$, D. Rau$^{1}$, H. R. Ott$^{1}$, 
S. M. Kazakov$^{1}$, J. Karpinski$^{1}$, S. Wessel$^2$}

\affiliation{ 
$^{1}$ Laboratorium f\"{u}r Festk\"{o}rperphysik,
ETH-H\"{o}nggerberg, CH-8093~Z\"{u}rich, Switzerland \\
$^2$ Institut f\"{u}r Theoretische Physik,
ETH-H\"{o}nggerberg, CH-8093~Z\"{u}rich, Switzerland
}


\begin{abstract}

We report the results of measurements of the dc magnetic susceptibility $\chi$ 
and of the $^{23}$Na nuclear magnetic resonance (NMR) response of NaVGe$_{2}$O$_{6}$, a
material in which the V ions form a network of interacting one-dimensional spin $S=1$ chains.
The experiments were made at temperatures between 2.5 and 300 K. 
The $\chi(T)$ data suggest that the formation of the expected
low-temperature Haldane phase is intercepted
by an antiferromagnetic phase transition at $T_N=18$ K.
The transition is also reflected in the $^{23}$Na NMR spectra and the
corresponding spin-lattice relaxation rate $T_1^{-1}(T)$.
In the ordered phase, $T_1^{-1}(T)$ decreases by orders of magnitude with 
decreasing temperature, indicating the formation of a gap
of the order of 12 K in the magnetic excitation spectrum.

\end{abstract}
\maketitle


\subsection{1. Introduction}

Quasi-one-dimensional (1D) spin systems have recently been
the subject of both theoretical and experimental investigations.
It was shown theoretically \cite{Haldane} that for chains of
antiferromagnetically coupled integer spins
the ground state is separated by an energy gap $\Delta_H$ from the magnetic excitations.
This was later confirmed experimentally for many systems \cite{Review1D},
including the inorganic compounds AgVP$_2$O$_6$ \cite{AgVP2S6},
CsNiCl$_3$ \cite{CsNiCl3}, YBaNiO$_5$ \cite{YBaNiO5},
as well as for organic substances \cite{Nonorganic1D}.

In 1999, a new member, LiVGe$_2$O$_6$, was added to the list of quasi-1D magnets
\cite{Millet1999}.
From the temperature dependence of the magnetic susceptibility,
the expected spin gap $\Delta_H/k_B$ was estimated to be about 19 K.
Instead of developing 
the Haldane phase at low temperatures, however, the compound exhibits a transition to a
three-dimensionally (3D) magnetically ordered state at 25 K.
This was established by NMR \cite{Gavi2000} \cite{Vonla2002},
as well as by neutron-diffraction experiments \cite{Lumsden2000}.

Since many aspects of the magnetically ordered state of LiVGe$_2$O$_6$ remain a
mystery, and the V-V interaction depends on details of the V-O-V bonds, we
consider it as worthwile to carefully investigate the magnetic properties of
other materials related to LiVGe$_2$O$_6$.
In this work, we report a study of the related compound NaVGe$_2$O$_6$.
Although LiVGe$_2$O$_6$ and NaVGe$_2$O$_6$ exhibit many similar features,
they are not identical at all. 
In particular, the differences in the ionic radii of Li and Na seem to be reflected in subtle structural changes. 
Our dc-susceptibility measurements indicate a magnetic phase transition at 18 K.
We provide a detailed analysis of the magnetic susceptibility of NaVGe$_2$O$_6$
(and also of LiVGe$_2$O$_6$), including computer-based model calculations. 
This leads to estimates of the intra- and interchain coupling parameters,
indicating that for NaVGe$_2$O$_6$, 
the intrachain coupling is weaker and the 1D-character less pronounced. 
From the estimated values we conclude that the non negligible
interchain coupling is the main cause for the absence 
of the expected formation of the Haldane state at low temperatures.
The results of subsequent $^{23}$Na NMR-experiments, 
including the mapping of NMR-spectra 
and measurements of both the spin-lattice and spin-spin relaxation rates,
confirm the existence of a phase transition, 
most likely to a 3D antiferromagnetically ordered state.

The article is organized as follows.
In Section 2 we provide information on the sample.
In Section 3 we report on the dc-susceptibility mesurements,
and in Section 4, 5 and 6 we describe and discuss the results of the NMR experiments.
In Section 7 we compare the magnetic properties 
of the related compounds NaVGe$_2$O$_6$ and LiVGe$_2$O$_6$.


\subsection{2. Crystal Structure and Sample}
\label{SecCrystStruct}

\begin{figure}[t]
 \begin{center}
  \leavevmode
  \epsfxsize=1.0\columnwidth \epsfbox{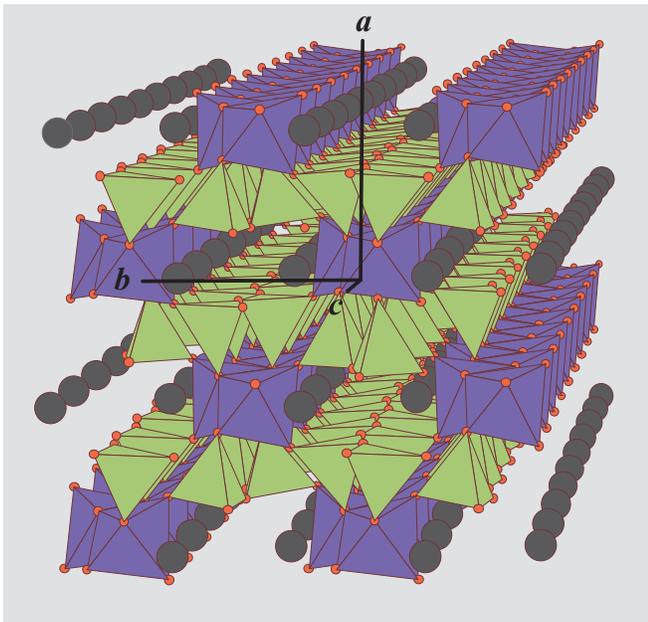}
\caption{
Schematic representation of the crystal structure of NaVGe$_{2}$O$_{6}$.
The crystalline axes are labelled.
The V$^{3+}$ ions are located in the centers of the dark-grey VO$_6$ octahedra,
while the Ge$^{4+}$ ions occupy the centers of the light-grey GeO$_4$
tetrahedra.
The Na$^+$ ions are represented by dark spheres.
}
\label{FigCrystStruct}
\end{center}
\end{figure}

NaVGe$_{2}$O$_{6}$ crystallizes with a monoclinic structure, space group
$P2_1/c$.
The NaVGe$_{2}$O$_{6}$ powder sample was prepared following the procedure
described in \cite{VasiVolo2002}.
A stoichiometric mixture of GeO$_2$ (Aldrich, 99.998$\%$),
V$_2$O$_3$ (Aldrich, 99.99$\%$),
and Na$_2$Ge$_2$O$_5$ was pressed into pellets and annealed at 900$^{\circ}$C for 70 h
in an evacuated silica tube.
Na$_2$Ge$_2$O$_5$ was synthesized by heating a mixture of
Na$_2$CO$_3$ (Aldrich, 99.995$\%$)
and GeO$_2$ at 760$^{\circ}$C for 15 h in air.
According to X-ray powder diffraction data,
the sample was of single phase with the pyroxene structure.
The corresponding lattice parameters are $a$=9.9600(5), $b$=8.8444(4), $c$=5.4858(2) 
\AA$\;$  and $\beta=106.50(2)^{\circ}$.
The crystal structure is shown schematically in Figure \ref{FigCrystStruct}.
It consists of chains of isolated, slightly distorted VO$_6$ octahedra joined at
the edges.
These chains are linked but also kept apart by double chains of distorted
GeO$_4$ tetrahedra.

Taking into account the most likely oxidation states of O$^{2-}$, Na$^{1+}$ and
Ge$^{4+}$, the V ions are expected to be trivalent.
The magnetic moments are thus due to the two $3d$ electrons localized at the 
V$^{3+}$ ions, which form an $S=1$ system.
Assuming that the interaction beween these moments is due to exchange interaction
mediated by the $2p$-electrons of the O ions, the intrachain coupling
$J$ is anticipated to be much larger than the interchain coupling
$J_{\perp}$ because the former involves only one O site,
whereas the latter is spread over two of them.

For our measurements we used a powdered sample,
with randomly oriented grains and a mass of 127 mg,
corresponding to 4.04$\times 10^{-4}$ mol.
The diameters of the powder grains were less than 0.1 mm.


\subsection{3. Magnetic Susceptibility}

\begin{figure}[t]
 \begin{center}
  \leavevmode
  \epsfxsize=0.8\columnwidth \epsfbox {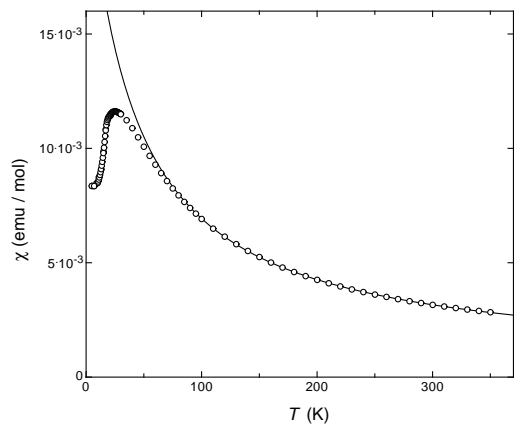}

\caption{
Magnetic susceptibility $\chi$ as a function of temperature $T$, 
measured in a magnetic field of $H=2$ T.
The solid line represents the best fit to the data for temperatures above
$100$ K, assuming $\chi(T)=\chi_0+C(T-\Theta)^{-1}$. }
\label{FigChiT}
\end{center}
\end{figure}

In order to confirm the above mentioned expected oxidation state of the V ions,
we measured the dc magnetic susceptibility $\chi(T)$ between 4 and 340 K, 
in various fixed magnetic fields between 0.01 and 2 T,
using a commercial SQUID magnetometer.
For temperatures above 100 K, the data can reasonably be fitted by a Curie-Weiss
type of law, i.e., $\chi(T)=\chi_0+C(T-\Theta)^{-1}$.
The constant term $\chi_0$ is attributed to magnetic background.
In Figure \ref{FigChiT} we display an example of $\chi(T)$, 
measured in a magnetic field of $H=2$ T.
In this field, $\chi_0=1.7\times 10^{-3}$ emu/mol.
The paramagnetic Curie temperature is $\Theta=-39\pm5$ K,
which signals the tendency of the V moments to couple antiferromagnetically.
The value of $C$ is
\begin{equation}
  C= 0.89 \pm 0.04 \;\mathrm{emu}\cdot\mathrm{mol}^{-1}\cdot\mathrm{K}
  \quad,
\end{equation}
where the indicated error range takes into account the values found for other
magnetic fields.
The effective magnetic moment (in units of $\mu_B$) is thus
\begin{equation}
  \label{Eqmueff}
  \mu_{\mathrm{eff}}=\sqrt{\frac{3k_B}{N\mu_B^2}C}=2.67\pm0.05
\end{equation} 
per V ion.
This value is close to the one expected for the V$^{3+}$ (and not V$^{4+}$) configuration,
with quenched orbital moments of the two $3d$ electrons \cite{AshMer}.

\begin{figure}[t]
 \begin{center}
  \leavevmode
  \epsfxsize=0.8\columnwidth \epsfbox {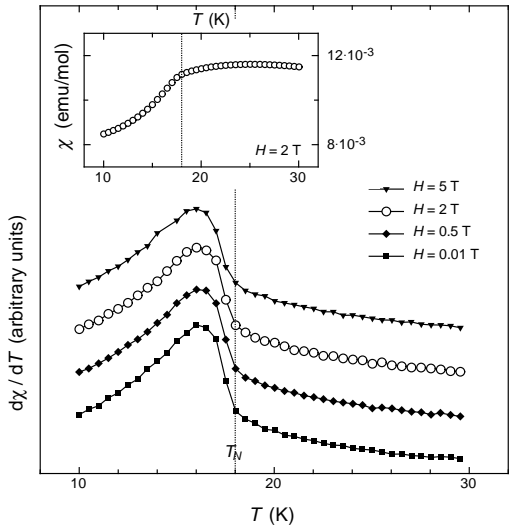}

\caption{
$d\chi/dT$ as a function of the temperature $T$ for different
applied fields. 
The inset shows $\chi(T)$ at $H=2$ T.
The dotted line indicates the low-field transition temperature $T_N=18$ K.
}
\label{FigdChidT}
\end{center}
\end{figure}

The maximum of $\chi(T)$ is reached at the temperature $T_{\mathrm{max}}= 25$ K.
Below this temperature, a kink in $\chi(T)$ is observed at $T_{\mathrm{N}}= 18$ K,
as demonstrated in the inset of Figure \ref{FigdChidT}.
The anomaly is much more evident by plotting $d\chi/dT(T)$.
The transition temperature $T_N$ 
is only little, if at all, affected by the value of the magnetic field up to 5
T.

\begin{figure}[t]
 \begin{center}
  \leavevmode
  \epsfxsize=0.8\columnwidth \epsfbox {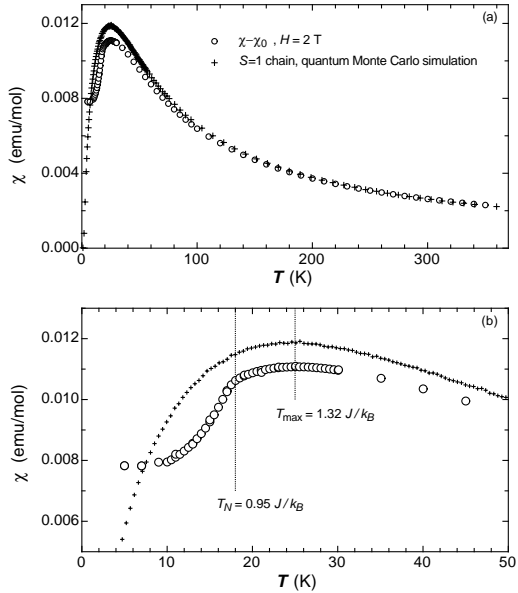}

\caption{
(a) Experimentally measured  $\chi(T)-\chi_0$ 
as a function of temperature in $H=2$ T,
and $\chi(T)$ resulting from quantum Monte Carlo calculations
for an $S=1$ spin chain.
Panel (b) emphasizes the range $T<50$ K. 
}
\label{FigChiSim}
\end{center}
\end{figure}

As already anticipated in Section 2,
the chemical composition and the crystal structure 
of NaVGe$_{2}$O$_{6}$ suggest that the V-moments form a quasi-1D $S=1$ spin system.
Nevertheless, the temperature dependence of the susceptibility
does not follow the expectations for a collection of independent $S=1$ spin chains.
This is evident if one compares the experimental data
with the results of quantum Monte Carlo simulations \cite{ALPS}, 
as shown in Figure \ref{FigChiSim}.
The simulation was performed for a single chain considering the Hamiltonian
\begin{equation}
  \label{EqChainHam}
  H=J\sum_i \vec{S}_i\cdot\vec{S}_{i+1}
  \quad,
\end{equation}
where $\vec{S}_i$ denotes a spin-1 operator at the $i$-th site,
and $J$ is the intrachain coupling.
$J$ was adjusted such that the maximum in $\chi(T)$ occurs at 25 K,
as is experimentally observed.
At high temperatures the experimental and calculated data coincide.
Near 25 K, however,
the experimental susceptibility turns out to be smaller than expected for the isolated
$S=1$-chains (see Figure \ref{FigChiSim}), suggesting the existence of some non negligible
interchain coupling.
The simulated and experimental $\chi(T)$ data
also differ significantly at low temperatures.
In particular, the continuous decrease to zero of the calculated $\chi(T)$,
revealing the influence of the Haldane gap, is not observed experimentally.
Instead a kink in $\chi(T)$ at 18 K indicates a magnetic phase transition.

As is well known from series expansions \cite{KogaKawa2000}, even a very small interchain coupling $J_{\perp}$
($J_{\perp}/J>0.026$) leads to the quenching of the Haldane gap
and induces three-dimensonal antiferromagnetic order
(see also \cite{KimBirg2000}).
As indicated by $\chi(T)$, this indeed seems to occur in
NaVGe$_{2}$O$_{6}$, suggesting a non-negligible interchain coupling $J_{\perp}$ in this compound.

\begin{figure}[t]
 \begin{center}
  \leavevmode
  \epsfxsize=0.8\columnwidth \epsfbox {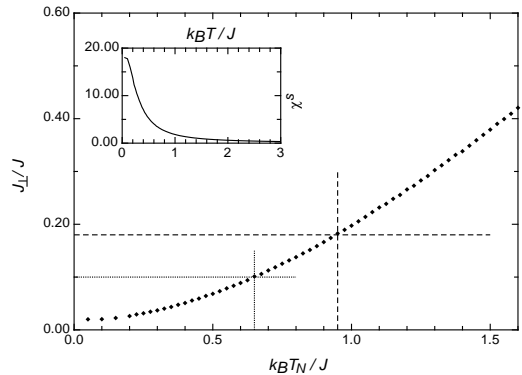}

\caption{
Interchain coupling as a function of the critical temperature,
according to equation (\ref{EqDefJperpTcrit}).
The extrapolation to $k_BT_N/J=0$ gives $J_{\perp}/J=0.02$.
The parameters $J_{\perp}/J$ and $k_BT_N/J$ for NaVGe$_{2}$O$_{6}$ are
indicated by the dashed lines, while the dotted lines represent
$J_{\perp}/J=0.1$ and the corresponding $k_BT_N/J$.
The inset shows the staggered spin susceptibility $\chi^s$ of a  
$S=1$ spin chain as a function of temperature, as calculated by quantum Monte Carlo
simulations.
}
\label{FigChiAFSim}
\end{center}
\end{figure}
From the measured values of $T_{\mathrm{max}}$ and $T_N$
we can estimate the magnitudes of both the intra- and interchain coupling, 
and thus their ratio, as follows:
Assuming that $T_{\mathrm{max}}$ is unaffected by a small interchain coupling,
quantum Monte Carlo simulations of an isolated chain provide a relation
between $J$ and $T_{\mathrm{max}}$
of the form
\begin{equation}
  \label{EqJTmax}
  \frac{J}{k_B\;T_{\mathrm{max}}} = \frac{1}{1.32}\pm0.02
  \quad.
\end{equation}
The value of the interchain coupling can be estimated from the transition
temperature $T_N$, using a random-phase approximation (RPA). The relevant
equation is
\begin{equation}
  \label{EqDefJperpTcrit}
  J_{\perp}/J=\frac{1}{z\chi^s[k_BT_N/J]}
  \quad,
\end{equation}
where $\chi^s$ denotes the staggered spin susceptibility of an isolated $S=1$
spin chain. 
Within conventional RPA \cite{RPA} $z=4$ equals the number of nearest neighbour chains.
Recently, it was found that for both the $S=1/2$ case and the classical Heisenberg model 
(corresponding to $S\to\infty$), an improved estimate of $J_{\perp}/J$ results by using a 
renormalized value of 
\begin{equation}
  z=2.78
\end{equation} 
for $J_{\perp}/J<0.2$ \cite{Yasu2003}. 
We applied this modified RPA also to the $S=1$ case.
The result for $(J_{\perp}/J)$ versus $[k_BT_N/J]$ is shown in Figure \ref{FigChiAFSim}, along with
an inset for the staggered susceptibility $\chi^s[k_BT/J]$ of an $S=1$ spin chain, 
as obtained from quantum Monte Carlo calculations.
The limit $[k_BT_N/J]\rightarrow0$ determines the critical value
$J_{\perp}/J=0.02$, i.e., the interchain coupling necessary to suppress the
Haldane gap, which is in reasonable agreement with the value estimated in \cite{KogaKawa2000}.
For $J_{\perp}/J=0.1$ we find $k_BT_N/J=0.65\pm0.02$ (see dotted lines in Figure
\ref{FigChiAFSim}), a value which is in good agreement with the result
$k_BT_N/J=0.65\pm0.01$ of an explicit quantum Monte Carlo calculation for interacting chains.

The dashed lines in Figure \ref{FigChiAFSim} represent the parameters
$J_{\perp}/J$ and $k_BT/J$ for NaVGe$_{2}$O$_{6}$. 
Using equations (\ref{EqJTmax}) and (\ref{EqDefJperpTcrit})
for $T_{\mathrm{max}}=25\pm0.5$ K and $T_N=18\pm0.5$ K, 
we obtain for NaVGe$_{2}$O$_{6}$
\begin{equation}
    J/k_B = 18.9\pm0.5\; \mathrm{K}
\end{equation}
and
\begin{equation}
    J_{\perp}/J = 0.18\pm0.01
  \quad,
\end{equation}
i.e.,
\begin{equation}
    J_{\perp}/k_B= 3.4\pm0.2 \; \mathrm{K}
  \quad.
\end{equation}


\subsection{4. NMR Spectra}

With our NMR experiments we probed the $^{23}$Na nuclei with a nuclear spin $I=\frac{3}{2}$, 
a gyromagnetic factor of $\gamma=7.0746\times10^{7}$s$^{-1}$T$^{-1}$
and a quadrupolar moment $|e|Q=0.108|e|\times10^{-24}$ cm$^2$,
where $e$ is the electron charge.

$^{23}$Na NMR spectra were obtained 
by monitoring the integrated spin-echo intensity
as a function of the irradiation frequency,
in a fixed magnetic field $H$
or, alternatively, as a function of the magnetic field,
at a fixed irradiation frequency.
The spin echo was generated with a two-pulse $\pi/2$-delay-$\pi$ spin-echo sequence,
irradiating a frequency window of about 20 kHz.

\begin{figure}[t]
 \begin{center}
  \leavevmode
  \epsfxsize=0.8\columnwidth \epsfbox {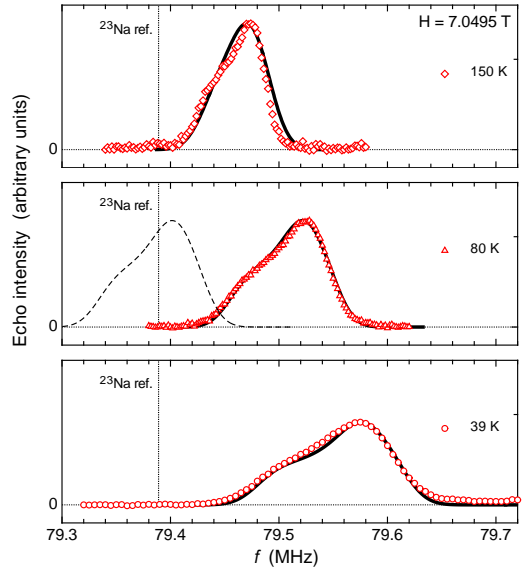}

\caption{
$^{23}$Na NMR spectra of the central line in the paramagnetic phase 
at three different temperatures in $H=7.0495$ T.
The integrated echo intensity (multiplied by the temperature $T$) is
represented as a function of the frequency $f$.
The solid lines represent the intensity $I(f)$, obtained using
Equation (\ref{EqIntensity}) and setting $A_0=1140$ G.
The dashed curve (shown only for $T=80$ K) corresponds to the calculation with
$A_0=0$ G.
The dotted line marks the reference position of the $^{23}$Na line (79.389 MHz).
}
\label{FigEIParam}
\end{center}
\end{figure}
In Figure \ref{FigEIParam}, 
we show examples of the central line of the spectra,
corresponding to the $\frac{1}{2}\leftrightarrow-\frac{1}{2}$ transition,
measured at three different temperatures above $T_N$,
in a fixed magnetic field of $H=7.0495$ T.
The background echo intensity, 
due to quadrupolar wings, i.e.,
the transitions $\pm\frac{3}{2}\leftrightarrow \pm\frac{1}{2}$, is subtracted.

\begin{figure}[t]
 \begin{center}
  \leavevmode
  \epsfxsize=0.8\columnwidth \epsfbox {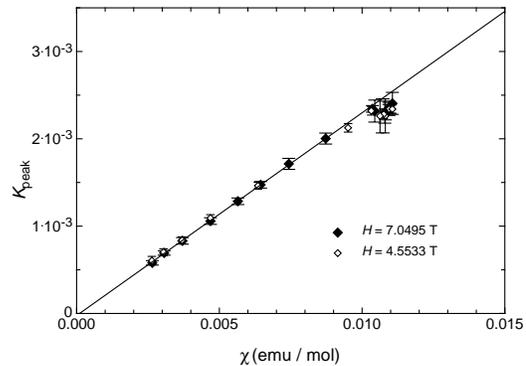}

\caption{
Relative frequency shift $K_{\mathrm{peak}}$ of the maximal intensity NMR line
as a function of the dc-susceptibility $\chi$, 
for temperatures above $T_{N}$.
The solid line is the best linear fit to the data at $H=7.0495$ T, above 60 K.}
\label{FigShiftChi}
\end{center}
\end{figure}
\begin{figure}[t]
 \begin{center}
  \leavevmode
  \epsfxsize=0.8\columnwidth \epsfbox {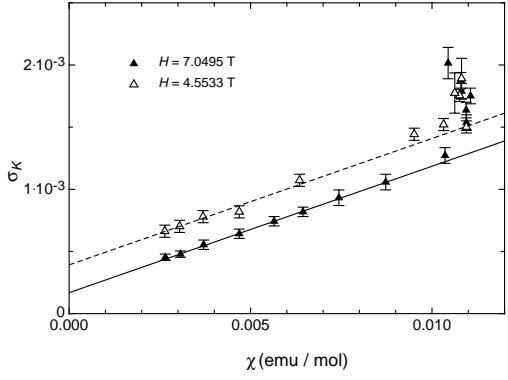}

\caption{
FWHM of the echo intensity of the central line,
divided by the $^{23}$Na reference frequency $f_0$,
as a function of the dc-susceptibility $\chi$, 
for temperatures above $T_{N}$.
The solid  and the dashed lines are the best linear fit to the data at
$H=7.0495$ T and $H=4.5533$ T respectively,
for temperatures above 60 K.}
\label{FigLinewidthChi}
\end{center}
\end{figure}

The paramagnetic character of the high-temperature phase is reflected
in Figures \ref{FigShiftChi} and \ref{FigLinewidthChi}.
In Figure \ref{FigShiftChi}, we plot the relative shift with respect to the
unshifted frequency $f_0$,
\begin{equation}
  K_{\mathrm{peak}}=\frac{f_{\mathrm{peak}}-f_0}{f_0}
  \quad,
\end{equation}
of the frequency $f_{\mathrm{peak}}$ with maximum echo intensity at different temperatures,
as a function of the Curie-Weiss part of the dc magnetic
susceptibility $\chi(T)$, measured at the corresponding temperatures
and in a magnetic field of $H=2$ T.
The data for the two different fields collapse onto the same curve.
Above $T_{\mathrm{max}}$,
$K(\chi)$ can reasonably well be fitted with a linear function,
indicating that the temperature dependence of the shift 
of the NMR-line at maximum intensity is of purely magnetic origin.
In Figure \ref{FigShiftChi}, we display the best fit to the data 
at $H=7.0495$ T and for $T>60$ K with the function
\begin{equation}
  K_{\mathrm{peak}}=a_{\mathrm{peak}}\chi+c_{\mathrm{peak}}
  \quad.
\end{equation}
Within experimental uncertainty, $c_{\mathrm{peak}}\approx0$.
The hyperfine coupling corresponding to the peak signal is \cite{Carter}
\begin{equation}
  \label{EqApeak}
  A_{\mathrm{peak}}=a_{\mathrm{peak}} \cdot N\mu_B=1300\pm 50 \;\mathrm{G}
  \quad.
\end{equation}

In Figure \ref{FigLinewidthChi}, we plot 
\begin{equation}
  \sigma_K=\frac{FWHM}{f_0}
  \quad,
\end{equation}
where $FWHM$ is the linewidth and $f_0$ is the unshifted frequency, 
at different temperatures,
as a function of the Curie-Weiss part of the dc-magnetic
susceptibility $\chi(T)$, which was measured at the corresponding temperatures 
and in a magnetic field of $H=2$ T.
For $T>T_{\mathrm{max}}$ a linear relation $\sigma_K(\chi)$, i.e.,
\begin{equation}
  \sigma_K=a_{\sigma}\chi+c_{\sigma}
  \quad,
\end{equation}
is observed.
In Figure \ref{FigLinewidthChi}, the solid line represents the best fit to the data 
for $H=7.0495$ T and for $T>60$ K.
The corresponding anisotropic hyperfine coupling is \cite{Carter}
\begin{equation}
  \sigma_A=a_{\sigma} \cdot N\mu_B=560\pm 50 \;\mathrm{G}
  \quad.
\end{equation}
From the value of $c_{\sigma}$ for $H=7.0495$ T we find a frequency broadening
$ \Delta f=c_{\sigma}f_0=14 \pm 2$ kHz,
which is attributed to the nonvanishing irradiation width that indeed is of
the same order of magnitude.
The dashed line is the best fit to the data for $H=4.5533$ T, keeping $a_{\sigma}$
fixed, resulting in $ \Delta f=c_{\sigma}f_0=18 \pm 2$ kHz, again of the same
order of magnitude.

The linear relation between $\sigma_K$ and $\chi$ breaks down for $T\leq T_{\mathrm{max}}$,
indicating that below that temperature a substantial degree 
of antiferromagnetic correlations interferes.

\begin{figure}[t]
 \begin{center}
  \leavevmode
  \epsfxsize=0.8\columnwidth \epsfbox {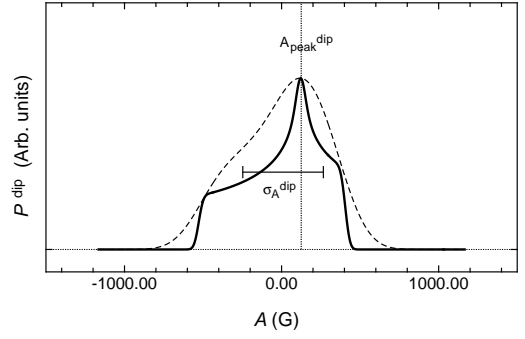}

\caption{
The solid line represents the shape $P^{\mathrm{dip}}$ of the central line 
calculated assuming dipolar coupling between
V moments and Na nuclei, as a function of the hyperfine coupling $A$.
The peak value $A_{\mathrm{peak}}^{\mathrm{dip}}$ and the estimated width
$\sigma_A^{\mathrm{dip}}$ are indicated.
The dashed line is obtained by broadening the solid line with $\Delta A=132$ G
(corresponding to the same brodaning $\Delta f$ in frequency space as used for
$T=80$ K in Figure \ref{FigEIParam}).
}
\label{FigDipolarLine}
\end{center}
\end{figure}

To complete the analysis of the features of the central line in the
paramagnetic phase, we compared the experimentally determined lineshape 
with the results of simulations.
They were performed by first calculating the lineshape $P^{\mathrm{dip}}(A)$
expected for a purely dipolar coupling between the V moments and the Na
nuclei, i.e., according to the formula
\begin{equation}
  \label{EqPdip}
  P^{\mathrm{dip}}(A)=
    \int_0^{2\pi}d\phi\int_0^{\pi}d\theta\;\sin\theta
    \;\delta\left(A-A^{\mathrm{dip}}(\theta,\phi)\right)
  \quad.
\end{equation}
Here, $\theta$ and $\phi$ are the spherical angles describing the orientation 
of the V moments of 1 $\mu_B$ with respect to the cristalline $c$-axis, 
$A^{\mathrm{dip}}(\theta,\phi)$ is the component of the induced
hyperfine-field at a Na site parallel to the V moments
and $\delta$ is the delta-function.
Equation (\ref{EqPdip}) implicitely assumes that
(\textit{a}) the powder sample consists of randomly oriented grains;
(\textit{b}) the V moments align along the applied external magnetic field.

The integral in Equation (\ref{EqPdip}) was calculated by approximating it with
the sum over a 200$\times$200-points-lattice $\mathcal{G}\subset[0,\pi]\times[0,2\pi]$,
\begin{equation}
  P^{\mathrm{dip}}(A)\sim
    \sum_{(\theta,\phi)\in\mathcal{G}}\sin\theta
    \;\Gamma_{\Delta A}\left(A-A^{\mathrm{dip}}(\theta,\phi)\right)
  \quad,
\end{equation}
where $\Gamma_{\Delta A}$ is a Gaussian function of width chosen as $\Delta
A\approx20$ G. 
The resulting lineshape $P^{\mathrm{dip}}(A)$ is represented as a function of 
the hyperfine coupling $A$ in Figure \ref{FigDipolarLine}.
The peak value is reached at
\begin{equation}
  A_{\mathrm{peak}}^{\mathrm{dip}}=125 \pm 10 \;\mathrm{G}
  \quad,
\end{equation}
while the mean square deviation is
\begin{equation}
  \sigma_A^{\mathrm{dip}}=520 \;\mathrm{G}
  \quad.
\end{equation}
While $\sigma_A^{\mathrm{dip}}$ is in good agreement with the experimental
value $\sigma_A$, $A_{\mathrm{peak}}^{\mathrm{dip}}$ is a factor
of ten smaller than $A_{\mathrm{peak}}$ (see Eq.(\ref{EqApeak})), indicating an additional, relevant non-dipolar
isotropic hyperfine-field coupling.

For a fixed temperature $T$ and the corresponding susceptibility $\chi$,
the signal intensity $I(f)$  as a function of the irradiation frequency $f$
was obtained by
\begin{equation}
  \label{EqIntensity}
  I(f)= I_{0,T}\cdot\left[ P^{\mathrm{dip}}(A(f)-A_0) \right]_{\Delta f}
  \quad,
\end{equation}
where $A$ and $f$ are related by
\begin{equation}
  \frac{f-f_0}{f_0}=\frac{A}{N\mu_B}\chi
  \quad,
\end{equation}
$I_{0,T}$ is a (temperature dependent) prefactor,
and $[\ldots]_{\Delta f}$ indicates the broadening in frequency space with a Gaussian
function of width $\Delta f$.

In Figure \ref{FigEIParam} we plot the calculated lineshapes with $\Delta
f=15\pm1$ kHz, $A_0=1140$ G,
and $I_{0,T}$ such that the experimental intensity at the frequency
$f_{\mathrm{peak}}$ is matched.
The parameter $A_0$ takes into account the above mentioned
non-dipolar coupling, 
whose value has been adjusted to reproduce the experimental data, i.e.,
\begin{equation}
  A_{\mathrm{peak}}\approx A_{\mathrm{peak}}^{\mathrm{dip}}+A_0
  \quad.
\end{equation}
The experimental lineshape is clearly very well accounted for by the
simulations.
The deformation range of the central NMR-line due to second order quadrupolar
effects is given by \cite{Carter}
\begin{equation}
  \Delta f_Q^{(2)}=\frac{\nu_Q}{144f_c}
                   \left[I(I+1)-\frac{3}{4}\right]
                   (\eta^2+22\eta+25)
  \quad,
\end{equation}
where $f_c$ is the Larmor frequency of the $^{23}$Na central line,
$\nu_Q=e^2qQ/2h$ is the quadrupolar frequency of $^{23}$Na
in the electric field gradient $eq$ and $\eta$ is the asymmetry parameter
of the electric field gradient
($eq=V_{zz}$ is the largest eigenvalue of the electric field gradient tensor,
and $\eta=(V_{xx}-V_{yy})/V_{zz}$).
As we shall see below, $\nu_Q=1.25$ MHz and $\eta=0.4$. 
With $f\sim79.6$ MHz this results in $\Delta f_Q^{(2)}\sim14$ kHz.
The effect is more evident at high temperatures, because it is less masked by magnetic
broadening. 
The small deviations of the order of 10 kHz, observed in particular at 150 K,
are attributed to such quadrupolar effects.

\begin{figure}[t]
 \begin{center}
  \leavevmode
  \epsfxsize=0.8\columnwidth \epsfbox {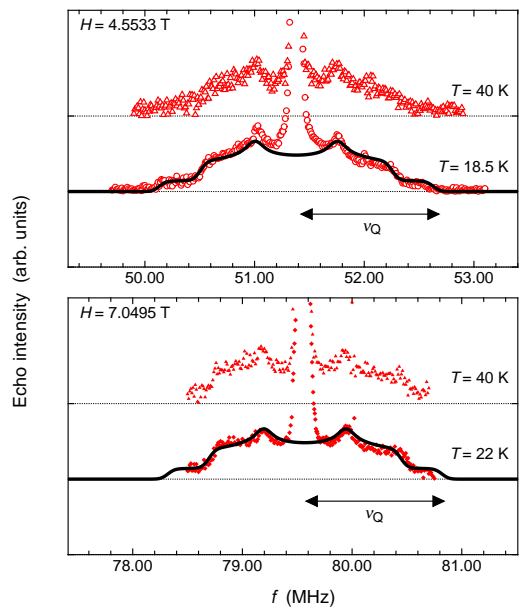}

\caption{
$^{23}$Na NMR spectra around the central line
in NaVGe$_{2}$O$_{6}$ at two different temperatures above $T_N$, and in two different fields.
The solid lines represent the calculated lineshapes due to quadrupolar effects.
}
\label{FigQuadWings}
\end{center}
\end{figure}

As shown in Figure \ref{FigQuadWings}, with decreasing temperatures but still above $T_N$,
we identify features near the central line which are temperature and field independent.
They are the signals of the wings due to the transitions $\pm\frac{3}{2}\leftrightarrow \pm\frac{1}{2}$.
This claim is supported by the results of comparisons 
of the measured signals with the signal calculated assuming that the intrinsic part of the central line, 
at frequency $f_c$, is unaffected by the orientation of the grains.
The shift of the $\pm\frac{3}{2}\leftrightarrow \pm\frac{1}{2}$-transitions
is given by \cite{Abragam}
\begin{equation}
  \Delta f_Q^{(1)}=\frac{\nu_Q}{2}
                   \left[3\cos^2\theta-1-\eta\sin^2\theta\cos2\phi\right]
  \quad.
\end{equation}
Here, $\theta$ and $\phi$ are the spherical angles 
describing the orientation of the principal axes of the electric field
gradient tensor with respect to the direction of the applied magnetic field.
The solid lines in Figure \ref{FigQuadWings} represent the calculated signal
without the contribution of the central line.
The best coincidence is achieved with a
quadrupolar frequency is $\nu_Q= 1.25$ MHz and an asymmetry parameter $\eta=0.4$.
The central frequencies $f_c$ were chosen to coincide with the values of the average frequencies of the
central line at the corresponding temperature, i.e., to
$51.38$ MHz (at 4.5533 T) and $79.57$ MHz (at 7.0495 T), respectively.
The largest eigenvalue of the electric field gradient tensor at the Na-sites is
\begin{equation}
  V_{zz}=eq=\frac{2h\nu_Q}{eQ}=9.58\cdot10^{20}\mathrm{V/m}^2
  \quad.
\end{equation}

\begin{figure}[t]
 \begin{center}
  \leavevmode
  \epsfxsize=0.8\columnwidth \epsfbox {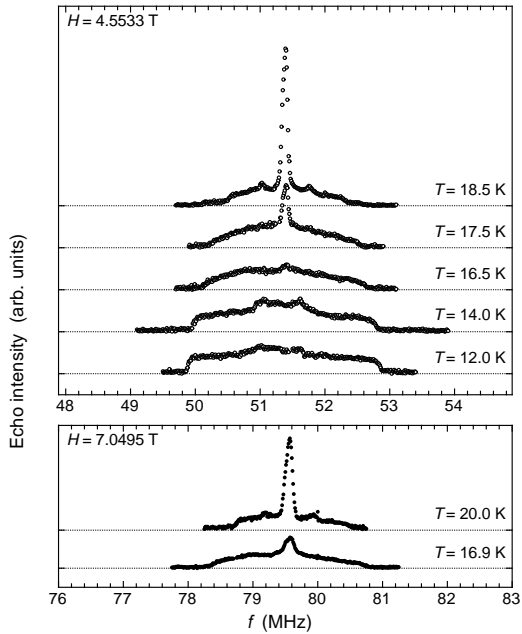}

\caption{
$^{23}$Na NMR spectra of NaVGe$_{2}$O$_{6}$.
The echo intensity is  represented as a function of the frequency $f$ for 
selected temperatures below 20 K.
The measurements were performed in fixed magnetic fields 
$H=4.5533$ T (upper panel) and $H= 7.0495$ T (lower panel).
}
\label{FigEIPhTr}
\end{center}
\end{figure}
Next, we focus our attention on the $^{23}$Na NMR spectra around and 
below the critical temperature $T_N=18$ K.
In Figure \ref{FigEIPhTr}, we show examples of such spectra, measured in the 
temperature range between $12$ K and $20$ K,
in fixed magnetic fields $H=4.5533$ T and $H=7.0495$ T, respectively.

\begin{figure}[t]
 \begin{center}
  \leavevmode
  \epsfxsize=0.8\columnwidth \epsfbox {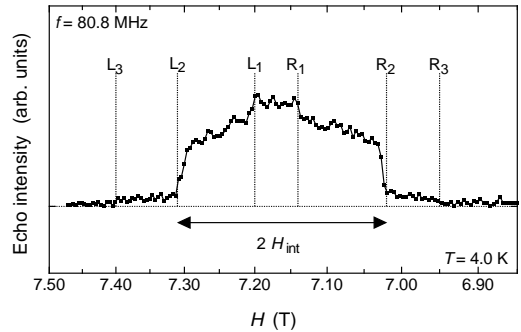}

\caption{
Example of a $^{23}$Na NMR spectrum of NaVGe$_{2}$O$_{6}$ at 4 K.
The echo intensity is plotted as a function of the magnetic field $H$.
The measurements were performed at a fixed irradiation frequency of 80.8 MHz.
}
\label{FigEIPhTrHSw}
\end{center}
\end{figure}

The phase transition at $T_N=18$ K is reflected by a drastic change of the
lineshape.
We note that the central-line intensity decreases rapidly between 18.5 K and 16.5 K,
indicating that the paramagnetic phase disappears upon crossing the
critical temperature from above.
At the same time, a much broader signal develops below 18.5 K.
At 14.5 K the signal has a well defined, ``nearly rectangular'' shape, 
which remains unchanged down to at least 4 K,
as evidenced by the signal obtained from a field sweep 
at a fixed frequency of 80.8 MHz, which is shown in Figure \ref{FigEIPhTrHSw}.
The observed width and the evolution of the lineshape below $T_N$ 
are both similar at different values of the magnetic field,
suggesting that the broadening of the line is due to ordered magnetic moments.
This allows, in particular, to exclude the formation of the Haldane phase
below $T_N$.
Indeed, our NMR results cannot be reconciled with the expectations
for a non-magnetic ground state, where a sharp narrowing of
the NMR line and a shift towards the reference frequency $f_0$
should be observed at temperatures
$T<\Delta_H/k_B\sim0.4\;J/k_B=7.6$ K \cite{YamaMiya1993}.

The deviation of the observed low-temperature lineshape from a rectangle,
which is expected for a purely magnetic broadening due to a fixed internal field,
in particular the feature in the range between the marks L$_1$ and R$_1$ in
Figure \ref{FigEIPhTrHSw}, 
is attributed to first order quadrupolar effects.
It is not a trivial task to predict the lineshape due to the combined effects
of an external field, an internal field and an electric field gradient.
The difficulty lies in the uncertainty of the directions of the ordered moments 
and of the principal axes of the the electric field gradient tensor at the
Na-sites, respectively.
Therefore, we discuss this aspect only qualitatively.
We assume that at low-temperature ($T\ll T_N$) the signal due to the 
$\frac{1}{2}\leftrightarrow-\frac{1}{2}$
transition extends from line L$_2$ to line R$_2$ in Figure \ref{FigEIPhTrHSw}.
The intensities in the ranges between L$_3$ and L$_2$, as well as between R$_2$ and R$_3$, 
reflect the quadrupolar wings.
Their extension, corresponding to about 1 MHz, is compatible with the
previously calculated quadrupolar frequency $\nu_Q$.
With these assumptions, the magnitude of the internal field,
indicated in Figure \ref{FigEIPhTrHSw},
is $H_{\mathrm{int}}=1470$ G.

\begin{table}[t]
\begin{center}
\begin{tabular}{|c|c|c|}
\hline
Direction of alignement & Interchain & $\phantom{oooo}H_{\mathrm{int}}\phantom{oooo}$ \\
of the moments          & ordering   & (G) $\pm$ 2 \% \\
\hline
$a$-axis & Ferromagnetic  & 1720  \\
         & Antiferromagnetic  & 570   \\
\hline
$b$-axis & Ferromagnetic  & 1550   \\
         & Antiferromagnetic  & 1050   \\
\hline
$c$-axis & Ferromagnetic  & 1720   \\
         & Antiferromagnetic  & 1720   \\
\hline
\end{tabular}
\caption{
Internal field $H_{\mathrm{int}}$ at the Na-sites,
for different orientations of the V-moments
and for either ferromagnetic or antiferromagnetic interchain ordering.
Ordered moments of $1.89\;\mu_B$ were assumed.
}
\label{TabHint}
\end{center}
\end{table}

As a plausibility check for the value of the internal field
$H_{\mathrm{int}}$ established above, we computed the dipolar field 
at the Na-sites due to three-dimensionally ordered magnetic moments at the V sites.
The $g$-factor for the V$^{3+}$ ions is given by
the effective moment found in equation
(\ref{Eqmueff}) as $g=\mu_{\mathrm{eff}}/\sqrt{S(S+1)}=1.89$.
In Table \ref{TabHint}, we show the values of the calculated dipolar fields
which are identical for all the Na-sites.
They were calculated assuming ordered moments of $1.89\;\mu_B$, oriented
along different axes, and for either ferromagnetic or antiferromagnetic
interchain order.
The obtained values are of the same order of magnitude as the experimental result,
but they do not allow to extract rigorous information about the orientation
of the ordered moments,
except that antiferromagnetic order along the $a$-axis seems very unlikely.
The uncertainity is, in particular, due to the fact that in the calculations 
we considered only dipolar coupling.
In addition, the moments need not be aligned along the same direction,
and their magnitude might also be reduced
(as was suggested in the case of LiVGe$_2$O$_6$, see \cite{Vonla2002}).


\subsection{5. NMR spin-lattice relaxation rate}

\begin{figure}[t]
 \begin{center}
  \leavevmode
  \epsfxsize=0.8\columnwidth \epsfbox {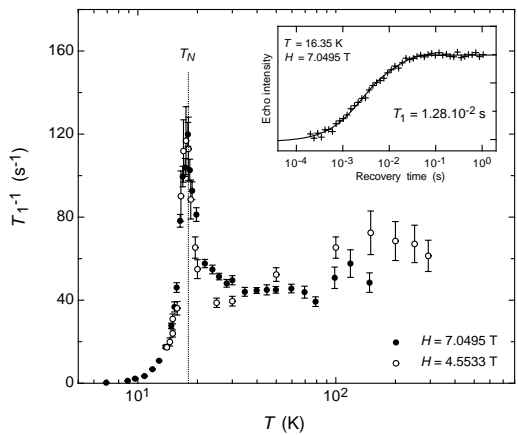}

\caption{
$^{23}$Na spin-lattice relaxation rate $T_{1}^{-1}$ as a 
function of temperature $T$.
The maximum of $T_{1}^{-1}$ at the critical temperature is indicated by the
dotted vertical line.
The inset shows an example of the echo intensity as a function of the recovery time; 
the solid line is the best fit using the recovery function (\ref{EqT1Rec}).
}
\label{FigT1}
\end{center}
\end{figure}

In order to probe the low-energy spin excitations in NaVGe$_{2}$O$_{6}$ at low
temperatures, we mesured the spin-lattice relaxation rate $T_1^{-1}(T)$ by 
monitoring the recovery of the $^{23}$Na nuclear magnetization after the 
application of a long comb of rf pulses, in the temperature range between $2$ K 
and $300$ K.
The experiments were performed by irradiating a frequency window of about 100 kHz.
Above $T_N$, the irradiation frequency was chosen to be 
in the center of the central line, while below $T_N$ we centered the window at approximatively
the center of the broadened line.
In both cases, a double recovery 
\begin{equation}
  \label{EqT1Rec}
  m(t)=m_{\infty}\left[1-D(0.6e^{-\frac{t}{6T_1}}+0.4e^{-\frac{t}{T_1}})\right]
  \quad,
\end{equation}
characteristic of  $I=3/2$ nuclei \cite{SiSuRo1962},
with $t$ the time variable and a constant $D\approx1$, was observed.

In Figure \ref{FigT1}, we display the temperature dependence of the spin-lattice 
relaxation rate.
A prominent peak in $T_1^{-1}(T)$ reflects the phase transition at $T_N=18$ K.
Above $T_N$, $T_1^{-1}$ varies only weakly with temperature
and is not much affected by changing the field from 4.5533 T to 7.0495 T.
Those features are typical for 
a spin-lattice relaxation driven by the spin-flips of the V$^{3+}$ 
paramagnetic moments whose dynamics is characterized 
by a short correlation time $\tau$, i.e., $\omega\tau\ll1$.
In this scenario \cite{Abragam}
\begin{equation}
  T_1^{-1}\sim \frac{2}{5}\gamma^2A^2S(S+1)\tau
  \quad,
\end{equation}
where $\gamma$ is the gyromagnetic ratio of the Na-nuclei,
$A$ is the hyperfine coupling and $S$ is the spin of the magnetic moments.
With $T_1^{-1}=50$ s$^{-1}$, $\gamma=7.0746\times10^{7}$s$^{-1}$T$^{-1}$ and
$A\sim0.1$ T, we estimate the correlation time of the V$^{3+}$-moments
to be $\tau\sim1.25\times10^{-12}$s.

\begin{figure}[t]
 \begin{center}
  \leavevmode
  \epsfxsize=0.8\columnwidth \epsfbox {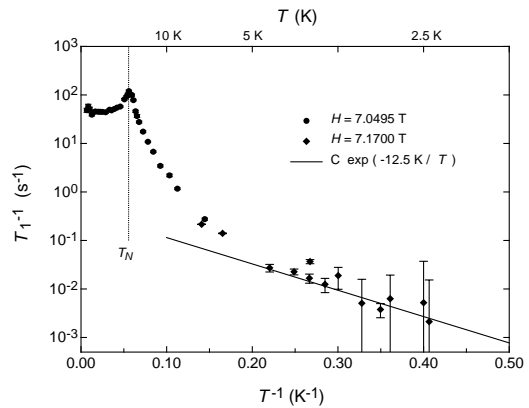}

\caption{
$^{23}$Na spin-lattice relaxation rate $T_{1}^{-1}$ as a 
function of inverse temperature $T^{-1}$.
The solid line represents the function 
$T_1^{-1}=C \exp(-\Delta/T)$ with $\Delta=12.5$ K.
}
\label{FigT1LowT}
\end{center}
\end{figure}

Below $T_N$, $T_1^{-1}$ decreases by orders of magnitude with decreasing 
temperature (a factor of $10^5$ from 18 K to 2.5 K, as shown in Figure \ref{FigT1LowT}), 
indicating the formation of a gap in the spectrum of magnon excitations in the
antiferromagnetically ordered state.
This is supported by $T_1^{-1}(T)$ well below $T_N$,
which varies according to
\begin{equation}
  \label{EqT1T}
  T_1^{-1}(T)\sim Ae^{-\frac{\Delta}{T}}
  \quad,
\end{equation} 
as is emphasized by the solid line in Figure \ref{FigT1LowT}.
The magnon processes which lead to spin-lattice relaxation
may be very complicated by involving several real and/or virtual magnons.
At low temperatures, however, it is fair to assume that a two-magnon process,
namely, the scattering of a magnon at the nucleus with a nuclear spin-flip
(Raman process) dominates \cite{Barak1974}.
In this case, $T_1^{-1}(T)$ is given by Equation (\ref{EqT1T}),
and $\Delta$ corresponds to the gap in the magnon spectrum.
Such a gap is usually attributed to an easy-axis single-ion magnetic anisotropy term in the
Hamiltonian, of the form $D\sum_iS_{i,z}S_{i,z}$ with $D<0$ \cite{JaccaMagnetism}.

In our case, the uncertainty of the data at the lowest temperatures
allows only for an estimate of the value of the gap $\Delta/k_B$,  
of the order of 12 K.
If a small amount of magnetic impurities, not indicated by the low-temperature
susceptibility, were present,
the actual value of the gap could be somewhat larger.

\subsection{6. NMR spin-spin relaxation rate}

\begin{figure}[t]
 \begin{center}
  \leavevmode
  \epsfxsize=0.8\columnwidth \epsfbox {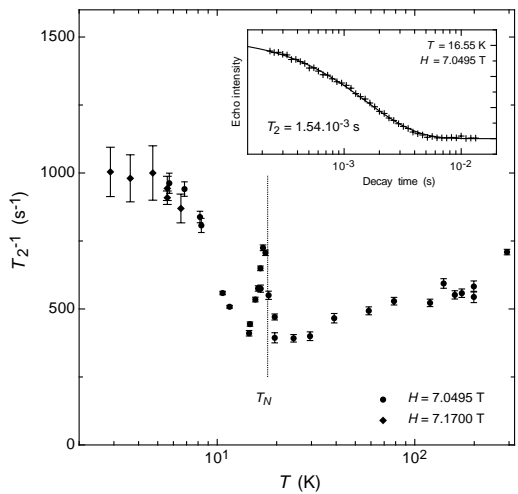}

\caption{
$^{23}$Na spin-spin relaxation rate $T_{2}^{-1}$ as a 
function of temperature $T$.
The inset shows an example of the echo intensity as a function of the decay time;
the solid line is the best fit using the exponential decay function in equation
(\ref{EqT2Decay}).
}
\label{FigT2}
\end{center}
\end{figure}

Our measurements of the spin-spin relaxation rate $T_{2}^{-1}$ 
as a function of temperature were made in the range between 2 K and 300 K.
The spin-echo lifetime $T_2^{\ast}$ was obtained 
by monitoring the spin-echo intensity as a function of the pulse delay $\tau$
in the $\pi/2$-delay-$\pi$ echo sequence.
The irradiation conditions were chosen to be the same as in the $T_1$-experiments.
At all temperatures, the intensity decayed according to
\begin{equation}
  \label{EqT2Decay}
  I(t)=I_0e^{-\frac{\tilde{t}}{T_2^{\ast}}}
  \quad.
\end{equation}
Here, $\tilde{t}=2(2/\pi\cdot t_{\pi/2}+d)+t_{\pi}\approx 2d$ \cite{Slichter}, where $t_{\pi/2}$ and
$t_{\pi}$ are the durations of the pulses, and $d$ is the delay between them.
The spin-spin lattice relaxation rate was then calculated from
\begin{equation}
  T_2^{-1}=T_2^{\ast-1}-T_1^{-1}\approx T_2^{\ast-1}
  \quad.
\end{equation}
The last approximation is justified because $T_2^{\ast-1}\approx10\cdot T_1^{-1}$.

In Figure \ref{FigT2}, we display the temperature dependence of the spin-spin 
relaxation rate, measured for $H=7.0495$ T.
A prominent peak in $T_2^{-1}(T)$ is observed at $17.5\pm0.5$ K,
and reflects the phase transition at $T_N$.
Below that peak,  $T_2^{-1}(T)$ exhibits an unexpected upturn with decreasing 
temperature and approaches a constant value at temperatures less than 5 K.
Generally, $T_2^{-1}$ is determined by the slow fluctuations 
of the internal magnetic field at the Na-sites.
Since in our case $T_2^{-1}\gg T_1^{-1}$, these fluctuations must
basically be given by spin-spin interactions between the Na-nuclei.
In the antiferromagnetically ordered state, besides the direct dipolar
interaction, 
also a magnon mediated interaction is expected to play a role.
At any rate, it is not easy to see how this would lead to an increase of
$T_2^{-1}$ at lower temperatures and we have no reasonable explanation for 
this particular feature.


\subsection{7. Comparison with LiVGe$_2$O$_6$}

From the structural point of view, LiVGe$_2$O$_6$ is very similar 
to NaVGe$_{2}$O$_{6}$.
More precisely, the two substances share the same space group $P2_1/c$, and 
the lattice parameters differ by less than 1.5\%.
Moreover, both compounds contain chains of VO$_6$-octahedra,
kept apart by GeO$_4$-tetrahedra \cite{Millet1999}. 
Previously, we and other authors reported dc-susceptibility and NMR studies
on LiVGe$_2$O$_6$ \cite{Millet1999}\cite{Gavi2000}\cite{Vonla2002}.
The results turned out to be qualitatively the same as those
for NaVGe$_{2}$O$_{6}$ reported in this work.
Below we focus our attention on the most evident quantitative differences.

The dc-susceptibility of LiVGe$_2$O$_6$ $\chi(T)$ 
shows a maximum at $T_{\mathrm{\max}}=62$ K and a kink at $T_N=25$ K,
from which the intra- and interchain couplings may be obtained in the manner
described in Section 3.
The calculated values are shown in Table \ref{TabLiNaSusc}.
While the interchain coupling turns out to be approximatively the same,
the intrachain coupling in LiVGe$_2$O$_6$ is about a factor 3 larger than 
that in NaVGe$_2$O$_6$.
This is somewhat surprising and indicates that even small differences in the
orientations of the orbitals involved in the exhange interaction can affect
the effective coupling between the moments at the V sites.

\begin{table}[t]
\begin{center}
\begin{tabular}{|l|c|c|}
\hline
           & LiVGe$_2$O$_6$ & NaVGe$_2$O$_6$ \\
\hline
$T_{\mathrm{max}}$ (K) & 62   & 25   \\
$J/k_B$ (K) & 47.0   & 18.9  \\
$T_N$ (K) & 25  & 18   \\
$J_{\perp}/k_B$ (K) & 3.4   & 3.4  \\
$J_{\perp}/J$ & 0.07 & 0.18 \\
\hline
\end{tabular}
\caption{
Comparison of data extracted from dc-susceptibility measurements
on LiVGe$_2$O$_6$ and NaVGe$_2$O$_6$.
}
\label{TabLiNaSusc}
\end{center}
\end{table}

LiVGe$_2$O$_6$ has been the subject not only of NMR-experiments,
but also of neutron scattering \cite{Lumsden2000}
and muon spin resonance experiments \cite{Blund2003}.
It has been concluded that the low-temperature phase
is antiferromagnetically ordered, with a ferromagnetic interchain coupling;
the magnitude of the ordered moments was reported to be $1.14\mu_B$,
a much smaller value than expected from the $g$-factor
inferred from the susceptibility.
Using the above value for the ordered moment and 
single crystal NMR-experiments results, the authors of \cite{Vonla2002}
suggested that the ordering occurs along the $c$-axis.
Our data are consistent with a similar scenario for NaVGe$_2$O$_6$.


\subsection{8. Conclusions}
NaVGe$_{2}$O$_{6}$ can be considered as a quasi 1-dimensional $S=1$ spin system
only at temperatures much higher than the corresponding equivalent of the
intrachain coupling $J/k_B\sim19$ K.
The expected Haldane phase, with a Haldane gap $\Delta_H/k_B$ of the order of 8 K,
does not develop because the system orders antiferromagnetically below 18 K.
The  interchain coupling, $J_{\perp}/k_B\sim 3.4$ K, estimated from the $\chi(T)$ data, 
is by far sufficient to explain the suppression of the Haldane gap.
The decay of the spin-lattice relaxation rate
$T_1^{-1}(T)\sim\exp(-\Delta/k_BT)$ observed at $T\ll T_N$,
indicates the opening of a gap $\Delta/k_B\leq 12$ K in the magnon excitation
spectrum of the three dimensional antiferromagnetically ordered system.
This phenomenon is attributed to fluctuations
of the two $3d$-electrons orbitals of the V$^{3+}$ ions, leading to
an easy-axis single-ion anisotropy term in the Hamiltonian.

Further experiments with single crystals,
aiming at obtaining more information on the orientation and the magnitude
of the components of the electric field tensor, the internal field 
and the eventual magnetic anisotropy seem to be in order.
 

\subsection{Acknowledgements}
We are grateful to M. Sigrist and F. Alet for useful discussions.
The numerical simulations were performed using the Asgard cluster at ETH Z\"urich.
SW acknowledges support by the Swiss National Funds.


\begin{thebibliography}{99}
\bibitem{ALPS}
The quantum Monte Carlo simulations were performed using a stochastic
series expansion \cite{Sandvik} code based on the ALPS library \cite{Troyer}.
\end{thebibliography}
\end{document}